\documentstyle[12pt]{article}
              \def\be{\begin{equation}}
              \def\ee{\end{equation}}
              \def\bea{\begin{eqnarray}}
              \def\eea{\end{eqnarray}}
        \def\ba{\begin{array}}
        \def\ea{\end{array}}

                  \def\d{\partial}
                  \def\s{\sum}
                  
                  \def\a{\alpha}
                  \def\eps{\epsilon}

                  \def\raw{\rightarrow}
                    \def\da{\dag}
                    \def\lm{\lambda}
                    
                    \def\D{\Delta}
                    \def\b{\beta}

                    \def\ah{\hat{A}}
                    \def\bh{\hat{B}}
                    \def\ch{\hat{C}}
                   \def\eh{\hat{D}}

\def\cs{\cdots}
\def\CN{{\cal N}}
\def\CM{{\cal M}}
\def\CA{{\cal A}}
\def\CB{{\cal B}}
\def\CD{{\cal D}}
\def\CO{{\cal O}}
\def\CD{{\cal D}}
\def\W{{\cal W}}
\def\lk{\left [}
\def\rk{\right ]}
\def\kt#1{\mid{{#1}}>}

\def\z#1{z_{#1}}
\def\de#1{\Delta_{#1}}
\def\ds{\Delta_\Psi}
\def\df{\Delta_\Phi}
\def\da{\Delta_A}
\def\db{\Delta_B}
\def\dc{\Delta_p}
\def\I{\rm {I\kern-.3em I}}
\def\C{\rm {I\kern-.520em C}}
\def\R{\rm {I\kern-.3em R}}
\def\CZ{\rm {Z\kern-.4em Z}}
\def\str{$sl(2,\C)$}
 
\def\unit{\rm {1\kern-.4em 1}}
\def\f{\frac}
        \def\ff{\f{1}{2}}
        \def\pp{<\Psi(\z1)\Psi(\z2)>}
\def\abf{<A(z_1)B(z_2)\Phi(z_3)>}
\def\abs{<A(z_1)B(z_2)\Psi(z_3)>}
\def\ab#1{\vert #1 \vert}
\def\kp{\kt{\de{p}}}
\def\kpp{\kt{{\D'}_p}}
\begin{document}
\begin{titlepage}
\vspace{-10mm}
\begin{flushright}
Feb. 1996\\
IPM-96-138\\
hep-th/9604007
\end{flushright}
\vspace{12pt}
\begin{center}
\begin{large}
             {\bf Logarithmic Operators in Conformal Field Theory
             and The $\W_\infty$-algebra}
\end{large}

\vspace{20pt}
{\bf A. Shafiekhani${}^a$\footnote{e-mail address: ashafie@theory.ipm.ac.ir}
and M.R. Rahimi Tabar${}^{a,b}$}

\vspace{20pt}
{\it ${^a}$Institute for Studies in Theoretical Physics and Mathematics\\
             P.O.Box: 19395-5531, Tehran, Iran\\
\vspace{6pt}
          ${^b}$   Dept. of Physics,
                 Iran University of Science and Technology,\\
                 Narmak, Tehran, Iran
}
\end{center}
\abstract{ It is shown explicitly that the correlation
functions of Conformal Field Theories (CFT) with the logarithmic 
operators are invariant under the differential realization of
Borel subalgebra of $\W_\infty$-algebra. This algebra is
constructed by tensor-operator algebra of differential representation of
ordinary \str.
This method allows us to write differential equations which can be used to
find general expression for three and four-point correlation functions 
possessing logarithmic operators. The operator product expansion (OPE) 
coefficients of general logarithmic CFT are given up to third level.
}
\vspace{1.0 cm}

\end{titlepage}
{\section {Introduction}}
It has been shown by Gurarie \cite{gu}, that in the OPE of two given local
fields which has at least two fields with the same conformal dimension, one
can find some operators with a special property, known as logarithmic
operators. As discussed in \cite{gu}, these operators together with the ordinary
operators form the basis of the Jordan cell for the operator $L_0$.
In some interesting physical theories,
for example dynamics
of polymers \cite{sa}, WZW model on the $GL(1,1)$ super-group \cite{srz},
percolation \cite{ga}, edge excitation in
fractional quantum Hall effect \cite{5}, one can find naturally
such operators.
Recently the role of such operators has been considered
in the study of some physical problems, for instance:
2D-magnetohydrodynamic turbulence
\cite{rr1,rr2,rr3},  2D-turbulence \cite{flo,rahim1},
$c_{p,1}$ models \cite{fl,cp1},
gravitationally dressed CFT`s \cite{bk,bk1} and in some critical disordered
models \cite{10,d14}. They also play an important role in so called unifying
$\W$ algebra \cite{W} and in the description of normalizable zero modes
for string backgrounds \cite{b16,b17}.

The basic properties of logarithmic operators are, that
they form part of the basis of the Jordan cell
for $L_0$ whose in the correlators posses logarithmic singularity.
It has been shown that in rational minimal models such a situation,
i.e. two fields with the same conformal dimensions, does not occur \cite{rr2}.
The modular invariant partition functions for such theories with $c_{eff}=1$
and the fusion rules of logarithmic conformal field theories (LCFT)
are considered in \cite{fl,fufl}.

In this paper, we consider the symmetry algebra
of the correlation functions with logarithmic factor.
We take the modified operator $L_0$ according to ref. [1] and show that
if certain conditions on the two point function of ordinary fileds with
logarthmic partner is satisfied, such correlation functions
remain invariant under $sl(2,\C)$ algebra.
Here, we introduce another method for calculating logarithmic operators.
This method is obtained by finding differential equations from
some combinations of ordinary $l_n$`s differential realization.
We show that this combination is related to the
Borel subalgebra of the $\W_\wedge$-algebra, which is
the wedge subalgebra of $\W_\infty$.

The structure of this paper is as follows: In section 2, we give a
summary of LCFT and its relation to $\W_\infty$-algebra.
In section 3, we calculate the general behavior
of three and four-point correlation functions of such theories. In
section 4, we give the OPE coefficients up to third level of such CFT's.
\vspace{0.5cm}
{\section {The Logarithmic Operators and $\W_\infty$-algebra}}

According to \cite{gu}, the OPE of two fields $A$ and $B$, which
their fusion rule \cite{bpz} contains two fields $\Phi$ and $\Psi$ of
equal dimension, has a logarithmic term,
\be\label{2op}
A(z,{\bar z})B(0,0)=\mid z\mid^{2(\df-\da-\db)}\left\{\Psi(0,0)+\cs+
\log {\mid z\mid}
[\Phi(0,0)+\cs]\right\},
\ee
where dots denote the descendants of fields $\Psi$ and $\Phi$.
To see this, it is sufficient to look at the four-point function \cite{bpz}:
\be\label{4p}
<A(\z1)B(\z2)A(\z3)B(\z4)>\sim\f{1}{(\z1-\z3)^{2\da}}\f{1}{(\z2-\z4)^{2\db}}
\f{1}{[x(1-x)]^{\da+\db-\df}}F(x)
\ee
where the cross ratio $x$ is given by:
\be
x=\f{(\z1-\z2)(\z3-\z4)}{(\z1-\z3)(\z2-\z4)}.
\ee
In degenerate models $F(x)$ satisfies a second order linear differential
equation \cite{bpz}.

It should be noted
that since logarithmic fields are not chiral, we can not separate holomorphic
and anti-holomorphic parts. In this paper $(z_i-z_j)$ is set to
$|z_i-z_j|^2$, for simplicity.

According to \cite{rr2}, the hypergeometric equation governing the
correlator of the two fields in whose OPE, two other fields $\Psi$ and $\Phi$
with conformal dimension $\ds$ and $\ds+\eps$ appear, admits two solutions
\be
{}_2F_1(a,b,c,x)
\ee
\be
x^\eps{}_2F_1(a+\eps,b+\eps,c+2\eps,x)
\ee
where $a$, $b$ and $c$ are sums of conformal dimension. Clearly in the limit
of $\eps\raw 0$ these two solutions coincide. However, it can be shown that
\cite{11} another independent solution
exists which involves logarithms and can be generated by standard methods.
 Therefore, the following two independent solutions can be constructed
according to [1,10].
\be
\s b_nx^n+\log x\s a_nx^n.
\ee
Next, consistency of equation (\ref{2op}) and (\ref{4p}) requires that
\be\label{e7}
<A(\z1)B(\z2)\Psi(\z3)>=<A(\z1)B(\z2)\Phi(\z3)>\left\{\log\f{(\z1-\z2)}
{(\z1-\z3)(\z2-\z3)}+\lm\right\},
\ee
\be\label{e8}
<\Psi(z)\Psi(0)>=-\f{2}{z^{2\ds}}[\log z+\lm'],
\ee
\be
<\Psi(z)\Phi(0)>=\f{1}{z^{2\df}},
\ee
where $\lm$ and $\lm'$ are constants.
In this notation $\Psi$ is the logarithmic operator and $\Phi$ is the
ordinary one.
Let us now consider the action of $sl(2,\C)$ on the above correlators.
In the absence of logarithmic
operators (e.g., in the rational minimal models \cite{bpz,rr2})
the correlation function is invariant under the action of \str,
\be
\s^{N}_{i=1}\d_{z_i}<\Phi_1(\z1)\cs\Phi_N(z_N)>=0
\ee
\be
\s^{N}_{i=1}(z_i\d_{z_i}+\de{i})<\Phi_1(\z1)\cs\Phi_N(z_N)>=0
\ee
\be
\s^{N}_{i=1}(z_i^2\d_{z_i}+2z_i\de{i})<\Phi_1(\z1)\cs\Phi_N(z_N)>=0
\ee
Indeed according to \cite{bpz,az} the operators
\be\label{dr}
l_-=\d_{z},\hspace{0.5cm}
l_0=z\d_{z}+\Delta,\hspace{0.5cm}
l_+=z^2\d_{z}+2z\Delta
\ee
are the differential realization of $sl(2,\C)$, with $\d_z=\f{\d}{\d z}$.
When there are logarithmic operators in CFT theories, the
ordinary correlation functions, for instance ordinary two-point function,
should be replaced by eq. (\ref{e8}).
According to \cite{gu}, in the logarithmic conformal field theory, $l_0$
is given by the following modified representation:
\be
[ L_0 , A(z)] = (z\d_{z} + \Delta_A + \CD ) A(z),
\ee
where the operator $\CD$ is such that $\CD\Phi (z)\equiv 0$
and $\CD\Psi (z)\equiv\Phi$.
This can be constructed by taking the OPE of $T(z)$ with $\Phi(0)$ and with
$\Psi(0)$. That is to say
\be
T(z) \Psi (0) = \f{\Delta}{z^2} \Psi (0) + \f{\d_z \Psi (0)}{z}
+ \f{\Phi (0)}{z^2} + \cdots,
\ee
\be
T(z) \Phi(0) = \f{\Delta}{z^2} \Phi (0) + \f{\d_z \Phi (0)}{z}
+ \cdots.
\ee
If we expand $T(z)$ by Laurant series we can write \cite{gu},
\be
[l_n , \Psi(z)] = (z^{n+1} \partial_z + \Delta (n+1) z^n) \Psi(z) +
(n+1) z^n \Phi(z),
\ee
\be
[ l_n , \Phi(z)] = (z^{n+1} \partial_z + \Delta (n+1) z^n) \Phi(z)
\ee
which means that we can redefine $l_n$ to $L_n$ such that:
\be
[ L_n , A(z) ] = (z^{n+1} \partial_z + \Delta_A (n+1) z^n + (n+1) z^n \CD) A(z)
\ee
with the property $\CD\Psi\equiv\Phi$ and $\CD\Phi\equiv 0$.
For example, one can show that eq.(8) is invariant under modified
$L_0$ ( i.e. non-diagonal $L_0$).
\be
\ba{c}
L_0 < \Psi (z) \Psi (0) > = < [ L_0 , \Psi (z) \Psi (0) ] > \cr
= < [ (z \d_z +  \ds) \Psi (z) + \Phi (z) ] \Psi (0) > +
< \Psi (z) \ds \Psi (0) > + < \Psi (z) \Phi (0)>\cr
= ( z \d_z + 2 \ds) < \Psi (z) \Psi (0) > +  < \Psi (z) \Phi (0) >
+ < \Phi (z) \Psi (0) >.
\ea\ee
By using eqs.(8) and (9) we find that:
\be
L_0 < \Psi(z) \Psi(0) > = 0.
\ee
Conversely, the eq.(19) allows us to write  Ward identities (for n=-1, 0, 1 )
for the calculation of the correlation functions. For instance, for two-point
functions we have:
\be
L_{-1} < \Psi(z) \Psi(0) >= L_{-1} < \Psi(z) \Phi(0) >= L_{-1} < \Phi(z) \Phi(0) > = 0
\ee
\be
L_0 < \Psi(z) \Psi(0) >= L_0 < \Psi(z) \Phi(0) >= L_0  < \Phi(z) \Phi(0) > = 0
\ee
\be
L_+ < \Psi(z) \Psi(0) >= L_+ < \Psi(z) \Phi(0) >= L_+  < \Phi(z) \Phi(0) > = 0.
\ee
Using eq.(23) we find,
\be
( z \d_z + 2 \ds) < \Psi (z) \Psi (0) > + 2 < \Psi (z) \Phi (0) > = 0,
\ee
\be
( z \d_z + 2 \ds) < \Psi (z) \Phi (0) > +  < \Psi (z) \Phi (0) >  = 0,
\ee
\be
( z \d_z + 2 \ds) < \Phi (z) \Phi (0) > = 0,
\ee
which can be solved yeilding:
\be
<\Phi(z)\Phi(0)>=-\f{a}{z^{2\ds}},
\ee
\be
<\Psi(z)\Phi(0)>=\f{1}{z^{2\ds}}[2a \log z+a'],
\ee
\be
<\Psi(z)\Psi(0)>=\f{1}{z^{2\ds}}[2a \log^2 z -2a' \log z +a'']
\ee
where so far $a, a'$ and $a''$ are arbitry parameters.
One can show that eq.(24) leads to the following relation:
\be
<\Phi(z)\Phi(0)> = 0
\ee
or $a=0$ which has already been pointed out in [15].
This means that if we insist on having conformal invariance, the two-point
function of ordinary fields which have logarithmic partner must be zero.

For $a\neq 0$ we don't have the conformal invariance, however the
correlation functions are invariant under $L_+ $ and $L_-$, which is the
subalgebra of $sl(2,\C)$. The condition (31) seems to be very restrictive
on two-point correlation functions.
For example, in $c=-2$ theory, which belongs to LCFT`s of $c_{p,1}$
series (with p=2) [20], the requirment of the conformal invariance forced sets
the two-point functions of fields with logarithmic partner to zero.
In other words, the ordinary fields with logarithmic partners
do not propagate.

Here, we introduce another method for finding certain
differential equations which can be used to obtain the correlation functions
by means of ordinary $l_n`s$.
This approach is based on the simple observation that in applying
the differential operator $ ( z \d_z + 2 \ds) $ (the differential
representation of $l_0$) on eq.(8), we obtain:
\bea
l_0 < \Psi(z) \Psi(0) > = < [ l_0 , \Psi (z) \Psi (0) ] > \equiv
(z \d_z + 2 \ds ) < \Psi(z) \Psi(0) > = -2 z^{-2 \ds}.
\eea
which behaves as eq.(9).
Similary the action of ordinary $l_\pm$ are as follows:
\be
< [ l_{-1} , \Psi(z) \Psi(0) ] > \equiv(\d_z) < \Psi(z) \Psi(0) >=0
\ee
\be
< [ l_+ , \Psi(z) \Psi(0) ] >\equiv(z^2\d_z+2z\ds)< \Psi(z) \Psi(0) >
= -2 z^{-2 \Delta_\Psi +1}.
\ee
Now we consider the action of $l_0$ and $l_+$ on the
correlation of $<\Psi(z) \Psi(0)>$. This can be written in the following form:
$$
< [ l_0 , [ l_+ , \Psi(z) \Psi(0) ] ] > = (z^2 \d_z + 2 z \ds)
(z \d_z +2 \ds) < \Psi(z) \Psi(0) > =
$$
\be
(z^2 \d_z + 2 z \ds)(z\d_z +2 \ds)
(-\f{2}{z^{2\ds}}[\log z+\lm']) = 0
\ee
Indeed, this is the action of differential representation of $l_0$ and $l_+$
on the logarithmic two-point function.

At first glance, it seems that the logarithmic correlation function,
$\pp$ is invariant under the set of
$$
\{l_-,\;l_0^2,\; l_-l_0,\; l_+l_0,\; l_0l_-l_+,\; l_-^2l_+,\; l_+l_-l_+\}
$$
where the last three of them are observed using the fact that $l_-l_+$ times
such correlation functions behave as ordinary correlation functions in CFT.
Using the commutation relation of $l_0$, $l_+$ and $l_-$,
\be\label{slt}
[l_0,l_\pm]=\pm l_\pm\hspace{0.5cm}
[l_+,l_-]=-2l_0,
\ee
the last three members of Eq. (25) can be written in terms of the first four
which we call $(A, B, C, D)$ respectively.

Simple calculations show that, the following algebraic relations
hold:
\be\ba{c}
[A,B]=-2C-A,\hskip 1cm [B,C]=2AB+C, \hskip 1cm [C,D]=-2l_0^3-2l_0
{\cdots}.
\ea\ee
Explicit calculation shows that this algebra is not closed. In each step,
we need to add new operators to the algebra.

On the other hand, the action of $l_0$ on logarithmic correlation functions
behave as expected. However the two-point function is the exception. In the
 case of three-point and four-point correlation functions, the action of
differential representation of $l_0$ on correlation
functions which have logarithmic term, behave like ordinary
three-point and four-point functions of CFT. We will come to this point
in the next section.
\\
In the rest of this section we determine the connection between the above
algebra and the
infinite dimensional $\W_\infty$-algebra.
The algebraic commutation relation of $V^i$'s, generators of spin-$(i+2)$
$\W_\infty$-algebra, are as follow \cite{po}:
\be\label{walg}
[V^i_m,V^j_n]=\s_{l\geq0}g_{2l}^{ij}(m.n)V^{i+j-2l}_{m+n}+c_i(m)\delta^{i,j}
\delta_{m+n,0}
\ee
where $g^{ij}_{2l}(m,n)$ are structure constants of the algebra and $c_i(m)$
are central terms. $V^i_m$ denotes the $m$th Laurent mode of spin-$(i+2)$
current $V^i(z)$. In our case, this current can be constructed from the
logarithmic operator $\Psi$ \cite{b16}. For $\W_\infty$-algebra with spins
$2\leq s<\infty$, the indices $i$ and $j$ range from $0$ to $\infty$.
The structure constants and central terms can be determined by demanding that
eq. (\ref{walg}) be consistent with the Jacobi identities \cite{po,13},
\be
c_i(m)=m(m^2-1)(m^2-4)\cs(m^2-(i+2)^2)c_i
\ee
where $c_i$'s are central charges, and the structure constants take the form
\be
g^{ij}_l(m,n)=\f{1}{2(l+1)!}\Phi^{ij}_lN^{ij}_l(m,n)
\ee
where $N^{ij}_l(m,n)$ are given by
\be
N^{ij}_l(m,n)=\s^{l+1}_{k=0}(-1)^k(^{l+1}_{k})[i+1+m]_{l+1-k}[i+1-m]_k
[j+1+n]_k[j+1-n]_{l=1-k}
\ee
and $[a]_n=a(a-1)\cs(a-n+1)=\f{a!}{(a-n)!}$.
The functions $\Phi^{ij}_l$ are given by
\be
\Phi^{ij}_l={}_4F_3\left [ \ba{cc}\ba{c}-\ff,\;\f{3}{2},\;-\f{l}{2}-\ff,\;
-\f{l}{2}\cr
-i-\ff,\;-j-\ff,\;i+j-l+\f{5}{2}\ea&;1\ea\right ]
\ee
where ${}_4F_3(1)$ is a generalized hypergeometric function \cite{13}.
Now consider a "wedge" of generators $V^i_m$ for which $\ab{m}\leq (i+1)$.
It can be easily shown that the commutator of any two generators within a
wedge only involves generators within the same wedge. The resulting wedge
subalgebra of $\W_\infty$ which is known as $\W_\wedge$, has the feature that
it is anomaly free, the central terms vanish for all commutators. The
algebra $\W_\wedge$ can be realized by generalization of $sl(2)$,
subalgebra of the Virasoro algebra, generated by $l_+$, $l_-$ and $l_0$.
Let us consider the tensor-operator algebra of \str. The tensor operator
algebra may be constructed in the following way. The \str\, generators
satisfy the commutation relations of eqs. (\ref{slt}). The Casimir
operator may be written as \be
Q=l_0^2-\ff(l_+l_-+l_-l_+).
\ee
The three generators, $l_-$, $l_+$ and $l_0$ transform as the {\bf 3}
representations of \str.
Higher-tensor operators, $T^l_m$ with ($-l\leq m\leq l$), transforming in
general as the $(2l+1)$-dimensional representation of \str, are
constructed from appropriate polynomials of degree $l$ in the $l$'s.
According to \cite{13}, we start with
the highest-weight state $T^l_l=(l_+)^l$, and then construct the lower-weight
states  in the usual way by $(l-m)$ successive application of $L_-$,
\be
T^l_m=\f{1}{(-2l)_{l-m}}[l_-,[l_-,\cs[l_-,(l_+)^l]\cs]]
\ee
where $(a)_n=a(a+1)\cs(a+n-1)=\f{(a+n-1)!}{(a-1)!}$.
Subtilties arise when we determine the exact relation of the above
construction with the $\W_\infty$-algebra.
Acting upon a Hilbert space in which the quadratic Casimir takes on a
definite value $Q=\mu$, the operators $T^l_m$ close into an
infinite-dimensional algebra which is parametrized by $\mu$, and known as
${\cal T}(\mu)$.
Therefore, we would expect that the above construction for
$\W_\wedge$-algebra should coincide with ${\cal T}(\mu)$ for some
specific value of $\mu$. It has been shown in ref. \cite{po}, that the
wedge subalgebra $\W_\wedge$ contained in $\W_\infty$-algebra
is the \str\, operator algebra ${\cal T}(0)$,
specified by the value $\mu=0$ for the quadratic Casimir. However for
any value of the parameter $\mu$, we can allow the representation to be
infinite dimensional \cite{po}.
It can be easily checked that in our case the logarithmic correlation
functions are not invariant under all of $T^l_m$'s, but they are invariant
under $T^l_m$ with $-l\leq m\leq 0$.
Therefore the important result is that the logarithmic correlation functions
are invariant under the Borel subalgebra of $\W_\wedge$-algebra.

\vspace{0.5cm}
{\section {Correlation Functions with Logarithmic Behavior}}

In this section, we calculate the correlation functions that have
a logarithmic term. A straightforward way for calculating the correlation
functions is to use Ward identities \cite{aziz2}.
For simplicity, first we consider the correlation functions of
conformal fields where only two of them have two fields of equal dimension in
their OPE. This can be done by using the
corresponding symmetry algebra, which was found in the last section.
It can be shown that in this case all correlation functions behave as follows:
\be\label{npf}
l_0<A(z_1)B(z_2)\CO(z_3)\cs\CO(z_n)>=k_n<\CO(z_1)\cs\CO(z_n)>
\ee
where the OPE of $A(z_{n-1})B(z_n)$ is given by eq. (\ref{2op}) and
 $\CO(z_i)$ are ordinary operators.
For the calculation of the two point function of such a conformal field
theory, we have
\be\label{l2p}
l_0<\Psi(z_1)\Psi(z_2)>=k_2<\CO(z_1)\CO(z_2)>=k_2\f{1}{z_{12}^{2\D_\Psi}}
\ee
and the following Ward identities:
\be\ba{ccc}
l_0 l_0 <\Psi(z_1)\Psi(z_2)>=0,\;&l_-<\Psi(z_1)\Psi(z_2)>=0,\;&l_+ l_0 <\Psi(z_1)\Psi(z_2)>=0,\
\ea\ee
and so on, where $z_{ij}=z_i-z_j$. Now we substitute the differential
realization of $l_0$ and $l_{\pm}$ from (\ref{dr}) into the above
equations by means of the "co-product",
\be\ba{c}
[z_1\d_{z_1}+z_2\d_{z_2}+2\D_{\Psi}]<\Psi(z_1)\Psi(z_2)>=k_2<\CO(z_1)
\CO(z_2)>\cr
[(z_1\d_{z_1}+z_2\d_{z_2}+2\D_{\Psi})(z_1\d_{z-1}+z_2\d_{z_2}+2\D_{\Psi})]
<\Psi(z_1)\Psi(z_2)>=0\cr
[\d_{z_1}+\d_{z_2}]<\Psi(z_1)\Psi(z_2)>=0\cr
[z_1^2\d_{z_1}+z_2^2\d_{z_2}+2(z_1+z_2)\D_{\Psi}]<\Psi(z_1)\Psi(z_2)>=0
\ea\ee
and so on. Solving the above equations for $<\Psi(z_1)\Psi(z_2)>$, we have
\be
<\Psi(z_1)\Psi(z_2)>=k_2<\CO(z_1)\CO(z_2)>\lk\log z_{12}+\lambda'\rk
\ee
where $k_2$ is $-2$ and $\lambda'$ is a constant.

A similar methods can be used for solving the three point correlation
function. Instead of (\ref{l2p}) we have
\be
l_0\abs=k_3\abf=
\f{k_3}{z_{12}^{\de{A}+\de{B}-\de{\Psi}}
z_{13}^{\de{A}-\de{B}+\de{\Psi}}z_{23}^{-\D_{A}+\D_{B}+\D_{\Psi}}}
\ee
and similar Ward identities for the three-point function. Solving these
equations gives:
\be
\abs=k_3\abf(a\log z_{12}+b\log z_{13}+c\log z_{23}+\lambda)
\ee
where $k_3, a, b$ and $c$ must satisfy the following relation:
\be
k_3+a+b+c=0.
\ee
For the four-point correlation function, $<A(z_1)B(z_2)\CO(z_3)\CO(z_4)>$, we
have
\be\ba{c}
L_0<A(z_1)B(z_2)\CO_1(z_3)\CO_2(z_4)>=k_4<\CO(z_1)\CO(z_2)\CO(z_3)\CO(z_4)>\cr
=\f{k_4}{z_{12}^{\de{A}+\de{B}-\de{\CO_1}+\de{\CO_2}}
z_{13}^{\de{A}-\de{B}+\de{\CO_1}-\de{\CO_2}}
z_{23}^{-\de{A}+\de{B}+\de{\CO_1}-\de{\CO_2}}
z_{34}^{2\de{\CO_2}}}F(x).
\ea\ee
Solving the relevant equations by a similar method, we find
\be\label{aboo}\ba{c}
<A(z_1)B(z_2)\CO(z_3)\CO(z_4)>=k_4<\CO(z_1)\CO(z_2)\CO(z_3)\CO(z_4)>\cr
\times\lk a\log z_{12}+b\log z_{13}+c\log z_{14}+d\log z_{23}+e\log z_{24}+
f\log z_{34}+\lambda"\rk
\ea\ee
where $\lambda"$ is a constant and $k_4$, $a$, $b$, $c$, $d$, $e$ and $f$ must
satisfy:
\be
k_4+a+b+c+d+e+f=0.
\ee

It can be seen that eq. (\ref{aboo}), is valid for four-point function of
$<A(z_1)B(z_2)\CO(z_3)\Psi(z_4)>$ type.
However correlation function of $<\Psi(z_1)\Psi(z_2)\Psi(z_3)>$ type cannot
be calculated by this method \cite{amir}.

\vspace{0.5cm}
{\section {OPE Coefficients}}

\subsection{OPE of Ordinary Conformal Fields}

The most general expression for the operator product expansion of ordinary
conformal fields is \cite{bpz,rr2}:
\be\label{opo}
\Phi_n(z,\bar{z})\Phi_m(0,0)=\s_p\s_k
z^{\de{p}-\de{n}-\de{m}+\s k_i}\;\bar{z}^{\bar{\Delta}_p-\bar{\Delta}_{n}
-\bar{\Delta}_{m}+\s {\bar{k}}_i}\;C_{nm}^{p,\{k\},\{\bar{k}\}}\;
\Phi_p^{\{k\},\{\bar{k}\}}(0,0)
\ee
where the coefficients are
\be\label{ce}\ba{c}
C_{nm}^{p,\{k\},\{\bar{k}\}}=C^p_{nm}\beta_{nm}^{p,\{k\}}
\bar{\beta}_{nm}^{p,\{\bar{k}\}},\hspace{0.75cm}
\{k\}=\{k_1,\;k_2,\;\cs,\;k_n\}.
\ea\ee
This OPE is the first regular hypergeometric function $F(x)$.
 Note that we have $\Phi_i(0)\mid 0>=\mid \de{i}>$. Acting
 on $\mid 0>$, by eq. (\ref{opo}) we obtain:
\be\label{phd}
\Phi_n(z)\mid \de{m}>=\s_pC_{nm}^p\;z^{\de{p}-\de{n}-\de{m}}\;\phi_p(z)\mid
\de{p}>
\ee
\be\label{phd2}\ba{c}
\phi_p(z)=\s_k z^{\s k_i}\;\beta_{nm}^{p,\{k\}}L_{-k_1}\;\cs L_{-k_n}\cr
\mid z,\de{p}>=\phi_p(z)\mid\de{p}>.
\ea\ee
Expanding $\mid z,\de{p}>$ in terms of the complete basis $\mid N,\de{p}>$,
i.e.,
\be\label{zde}
\mid z,\de{p}>=\s_Nz^N\mid N, \de{p}>,
\ee
and applying $L_j$ on eq. (\ref{phd}), we obtain:
\be\label{lj}
L_j\mid N+j,\de{p}>=(\de{p}-\de{m}+j\de{n}+N)\mid N,\de{p}>.
\ee
We also had that
\be\label{lo}
L_0\kt{N,\de{p}}=(\dc+N)\kt{N,\de{p}}.
\ee
By solving the recursion relations we can find $\beta_{nm}^{p,\{k\}}$.

For the first level we have
\be
L_{1}\mid 1,\de{p}>=(\de{p}-\de{m}-\de{n})\mid \de{p}>
\ee
which results in
\be\label{l1}
\kt{1,\de{p}}=\a_1L_{-1}\kt{\de{p}};\hspace{0.5cm}
\a_1=\f{\de{p}-\de{m}+\de{n}}{2\de{p}}.
\ee
For the second level by means of eqs. (\ref{zde}) and (\ref{l1}) we have
\be\ba{c}
L_{1}\kt{2,\de{p}}=\ah\a_1L_{-1}\kp\cr
L_{2}\kt{2,\de{p}}=\bh\kp
\ea\ee
which results in:
\be\label{l2}
\kt{2,\de{p}}=(\a_2L^2_{-1}+\a_3L_{-2})\kp
\ee
with $\a_2$ and $\a_3$ satisfying the following system of equations:
$$
{}_2\CN_{ij}\a_j={}_2\CA_i,\hspace{1cm} i,j=2,3
$$
where
\be\ba{c}
{}_2\CN=\lk\ba{cc} 4\de{p}+2&3\cr
6\de{p}&4\de{p}+\f{c}{2}\ea\rk
,\hspace{0.75cm}
{}_2\CA=\lk\ba{c}\ah\a_1\cr
\bh\ea\rk.
\ea\ee
and index $"2"$ refers to the level. This system can now be solved to give:
\be
\lk\ba{c}\a_2\cr \a_3\ea\rk=\f{1}{\de{1}}\lk\ba{c}
\ch-3\bh
\cr
-6\dc\ah\a_1+\eh\bh \ea\rk
\ee
where
\bea
\ah&=&\de{p}-\de{m}+\de{n}+1\cr
\bh&=&\de{p}-\de{m}+2\de{n}\cr
\ch&=&4\de{p}+\f{c}{2}\cr
\eh&=&4\de{p}+2\cr
\de{1}&=&c(2\de{p}+1)+2\de{p}(8\de{p}-5).
\eea
A similar method will work for higher levels. For the level N, we have in
place of eq. (\ref{l2}), an expansion corresponding to the partition of
N. We then find a system of equations by successively applying $L_j$, and
finally the coefficients are derived. For the third level, using Eq.
(\ref{lj}), we have: \be\ba{l}
L_1\kt{3,\de{p}}=(\ah+1)(\a_2L_{-2}+\a_3L^2_{-1})\kp\cr
L_2\kt{3,\de{p}}=(\bh+1)\a_1L_{-1}\kp\cr
L_3\kt{3,\de{p}}=(\bh+\de{n})\kp
\ea\ee
and $\kt{3,\de{p}}$ is given by
\be
\kt{3,\de{p}}=(\a_4L^3_{-1}+\a_5L_{-1}L_{-2}+\a_6L_{-3})\kp
\ee
where $\a_4$, $\a_5$ and $\a_6$ satisfy the following system of equations:
$$
{}_3\CN_{ij}\a_j={}_3\CA_i,\hspace{1cm} i,j=4,5,6
$$
with
\be
{}_3\CN=\lk\ba{ccc}
18\dc+6&4\dc+\f{c}{2}
+9&5\cr 24\dc&16\dc+2c&6\dc+2c\cr 6\dc+6&2\dc+7&4
\ea\rk\hspace{0.5cm}
{}_3\CA=\lk\ba{c} (\bh+1)\a_1\cr\bh+\de{n}\cr(
\ah+1)(\a_2+\a_3)\ea\rk.
\ee
Solving of the above system we obtain:
\be\ba{c}
\a_4=\f{-1}{6\D_2}{\Big [}2\a_1(\bh+1)(6\de{p}^2+6c\de{p}-11\de{p}+3c)
+(\bh+\D_n)(6\de{p}+2c+1)\cr
-(\ah+1)(\a_2+\a_3)(24\de{p}^2+11c\de{p}-26\de{p}+c^2+8c){\Big]}
\ea\ee
\be\ba{c}
\a_5=\f{1}{\D_2}{\Big [}2\a_1(\bh+1)(3\de{p}^2+c\de{p}-5\de{p}+c)
+(\bh+\D_n)(7\de{p}-1)\cr
-2(\ah+1)(\a_2+\a_3)(9\de{p}^2+3c\de{p}-7\de{p}+c){\Big ]}
\ea\ee
\be\ba{c}
\a_6=\f{1}{2\D_2}{\Big [}-4(\bh+1)\a_1(4\de{p}^2+c\de{p}-\de{p}+c)
+(\bh+\D_n)(-4\de{p}^2+c\de{p}\cr
-20\de{p}+c+4)
+4(\ah+1)(\a_2+\a_3)(16\de{p}^2+2c\de{p}-10\de{p}+c){\Big ]}
\ea\ee
where
$$
\Delta_2=
c^2(\dc+1)-c(\dc^2+11\dc-2)-4\dc(3\dc^2-7\dc+2).
$$

{\subsection {OPE of Logarithmic Conformal Fields }
The second most general solution of $F(x)$, which implies new OPE
 for two conformal fields which have two fields $\phi$ and $\psi$, of equal
dimension in their fusion rule, eq. (\ref{2op}), is:
\be\ba{c}\label{opl}
\Phi_n(z,\bar{z})\Phi_m(0,0)=\s_p\s_k
z^{\de{p}-\de{n}-\de{m}+\s k_i}\;\bar{z}^{\bar{\Delta}_p-\bar{\Delta}_{n}
-\bar{\Delta}_{m}+\s {\bar{k}}_i}\;\cr
\hspace{1cm}\times\lk C_{nm}^{p,\{k\},\{\bar{k}\}}
\log\ab{z}^2\phi_p^{\{k\},\{\bar{k}\}}(0,0)
+{C'}_{nm}^{p,\{k\},\{\bar{k}\}}\;\psi_p^{\{k\},\{\bar{k}\}}(0,0)\rk
\ea\ee
where $\phi_p^{\{k\},\{\bar{k}\}}$ is an ordinary conformal fields, which has
been discussed in the last subsection and $\psi_p^{\{k\},\{\bar{k}\}}$
denote the new pseudo-operator with unusual properties.

The coefficients $C_{nm}^{p,\{k\},\{\bar{k}\}}$ are the same as
eq.(\ref{ce}) and
\be\ba{c}
{C'}_{nm}^{p,\{k\},\{\bar{k}\}}={C'}^p_{nm}\beta_{nm}^{p,\{k\}}
\bar{\beta}_{nm}^{p,\{\bar{k}\}}\hspace{0.75cm}
\{k\}=\{k_1,\;k_2,\;\cs,\;k_n\}.
\ea\ee
Similar to the expressions for $\phi_p^{\{k\},\{\bar{k}\}}$, we have the
following for $\psi_p^{\{k\},\{\bar{k}\}}$:
\be
\psi_i(0)\kt{0}=\kt{\D'_p},\hspace{0.25cm}\mid z,{\D'}_p>=\psi_p(z)\kt{\D'_p}
\ee
and instead of eq. (\ref{phd}), we have
\be
\phi_n(z)\kt{\de{m}}=\s_p z^{\de{p}-\de{n}-\de{m}}\;[\log z\;C_{nm}^p\;
\phi_p(z)+{C'}_{nm}^p\;\psi_p(z)]\kt{\de{p}}
\ee
where
\be
\kt{z,\D'_{p}}=\psi_p(z)\mid\de{p}>
\ee
and its expansion in terms of $z^N$ is
\be\label{zdel}
\mid z,{\D'}_p>=\s_Nz^N\mid N, {\D'}_p>.
\ee
Collecting the above expressions together, we obtain the
following relations \cite{gu,rr1,bpz}
\be\label{lj2}
L_j\kt{N+j,\de{p}}=(\de{p}-\de{m}+j\de{n}+N)\mid N,\de{p}>
\ee
\be\label{ljl}
L_j\kt{N+j,{\D'}_p}=\kt{N,\dc}+(\dc-\de{m}+j\de{n}+N)\kt{N,{\D'}_p}
\ee
\be
L_0\kt{N,\de{p}}=(N+\dc)\kt{N,\de{p}}.
\ee
\be\label{lol}
L_0\kt{N,{\D'}_p}=\kt{N,\dc}+(\dc+N)\kt{N,{\D'}_p}.
\ee
Now by using (\ref{lj2}-\ref{lol}), we can calculate the OPE coefficients of
two conformal operators, which have at least two operators with equal
dimension in their OPE.

In the same way as in the ordinary case, for the first level we have
\be\ba{c}
\kt{1,{\D'}_p}={\a'}_1L_{-1}\kt{\dc}+\b_1L_{-1}\kt{{\D'}_p}\cr
L_1\kt{1,{\D'}_p}=(2\dc{\a'}_1+2\b_1)\kt{\dc}+2\dc\b_1\kt{{\D'}_p}\cr
=\kp+(\dc-\de{m}+\de{n})\kpp
\ea\ee
with ${\a'}_1$ and $\b_1$ satisfying the following system of equations;
\be
{}_1\CM_{ij}\gamma_j={}_1\CB_i;\hspace{0.5cm} i,j=1,2;
\hspace{0.5cm} \gamma_1=\b_1,\hspace{0.15cm}\gamma_2={\a'}_1.
\ee
where
\be\label{mb1}
{}_1\CM=\left [\begin {array}{cc} 2\dc&2\\\noalign{\medskip}0&2
\dc\end {array}\right ]\hspace{1cm}
{}_1\CB=\left[\begin{array}{c}1\cr(\ah-1)\end{array}\right]
\ee
which results in:
\be
\lk\ba{c}\b_1\cr{\a'}_1\ea\rk=\f{1}{2\dc^2}
\left [\begin{array}{c}{\de{m}-\de{n}}\cr \dc(\ah-1
)\end{array}\right].
\ee
It is not surprising that ${\a'}_1=\a_1$.

For the second level we have
\be
\kt{2,{\D'}_p}=({\a'}_2L^2_{-1}+{\a'}_3L_{-2})\kpp+
(\b_2L^2_{-1}+\b_3L_{-2})\kp.
\ee
Using (\ref{lj2}-\ref{lol}), we see that ${\a'}_2$, ${\a'}_3$, $\b_2$ and
$\b_3$ satisfy the following system of equations:
\be
{}_2\CM_{ij}\gamma_j={}_2\CB_i;\hspace{0.25cm} i,j=1,2,3,4;
\hspace{0.25cm} \gamma_1=\b_2,\hspace{0.15cm}\gamma_2=\b_3,\hspace{0.15cm}
\gamma_3={\a'}_2,\hspace{0.15cm} \gamma_4={\a'}_3
\ee
where
\be\label{mb2}
{}_2\CM=\left [\ba{cccc} 4\dc+2&3&4&0
\\\noalign{\medskip}6\dc&4\dc+\f{c}{2}&6&4
\\\noalign{\medskip}0&0&4\dc+2&3\\\noalign{\medskip}0&0&6
\dc&4\dc+\f{c}{2}\ea\rk
\hspace{0.5cm}
{}_2\CB=\lk\ba{c}
\ah\b_1+\a_1\cr 1 \cr
\ah\a_1\cr \bh\ea\rk.
\ee
These result in:
\be
\lk\ba{c}
\b_2\cr\b_3\cr{\a'}_2\cr{\a'}_3\ea\rk=
\f{1}{\D^2_1}\lk\ba{c}
\ch[\ah(\D_1\b_1-4\ch\a_1)+\D_1\a_1-24\bh]
+3[c(3\ah\a_1+2\bh)+10\bh-\D_1]\cr
\D_1(\eh-6\dc(\ah\b_1+\a_1))+6\ah\a_1(16\dc^2-c)
-4\bh(\eh-9)\cr
\ah\a_1\D_1\ch-3\bh\D_1\cr
-6\ah\a_1\D_1\dc+\bh\D_1(4\dc+2)\ea\rk.
\ee
For the third level, using eqs. (\ref{lj2}-\ref{lol}), we have:
\be
\kt{3,\de{p}}=(\b_4L^3_{-1}+\b_5L_{-1}L_{-2}+\b_6L_{-3})\kp
+(\a'_4L^3_{-1}+\a'_5L_{-1}L_{-2}+\a'_6L_{-3})\kpp
\ee
where $\a'_i$ and $\b_i$ ($i=4,\;5,\;6$) satisfy the following system of
equations:
\be
{}_3\CM_{ij}\gamma_j={}_3\CB_j,\;\;i,j=1,\cs,6,
\ee
$\gamma_i$'s ($i=1,2,3$) denote $\b_i$'s ($i=4,5,6$) and
$\gamma_i$'s ($i=4,5,6$) denote $\a'_i$'s ($i=4,5,6$), respectively.
The exact expression for ${}_3\CM_{i}$ and ${}_3\CB_{i}$ are:
\be
{}_3\CM=
\left [\begin {array}{cccccc} 18\dc+6&4\dc+9+{\frac
c{2}}&5&18&4&0\\\noalign{\medskip}24\dc&16\dc+2c
&6\dc+2c&24&16&6\\\noalign{\medskip}6\dc+6&2\dc
+7&4&6&2&0\\\noalign{\medskip}0&0&0&18\dc+6&4\dc
+9+{\frac c{2}}&5\\\noalign{\medskip}0&0&0&24\dc&16\dc
+2c&6\dc+2c\\\noalign{\medskip}0&0&0&6\dc
+6&2\dc+7&4\end {array}\right ]
\ee
\be
{}_3\CB=
\lk\ba{c} (\bh+1)\b_1+\a_1 \cr
1\cr (\ah+1)(\b_2+\b_3)+(\a_2+\a_3)\cr (\bh+1)\a'_1\cr
(\bh+\de{n})\cr (\ah+1)(\a'_2+\a'_3)\ea\rk.
\ee
In the above results, three significant remarks are inorder:\\
1- In the diagonal blocks of ${}_i\CM$, there are two copies of relevant
ordinary ${}_i\CN$.\\
2- The upper off diagonal block of ${}_i\CM$ is the derivative of the
diagonal block with respect to $\dc$ and the lower off-diagonal block is zero.

This is another evidence insupport of the fact that logaritmic operator together with ordinary operator
form the basis of Jordan cell for $L_0$.\\
Hence,
\be
{}_i\CM=\lk\ba{cc} {}_i\CN&\d_{\dc}\;{}_i\CN\cr
0&{}_i\CN\ea\rk \hspace{0.5cm}
{}_i\CB=\lk\ba{c}{}_i\CD\cr{}_i\CA\ea\rk
\ee
where $\d_{\dc}=\f{{\d}}{\d_{\dc}}$, ${}_i\CA$ is the same as in the last
subsection and ${}_i\CD$ is a column matrix with elements which resemble the
first half of ${}_i\CB$ in eqs. (\ref{mb1},\ref{mb2}).

Finally, $\b_i$s and $\a'_i$s are as follows:
\be\ba{c}
\lk\a'_i\rk=\lk\a_i\rk={}_i\CN^{-1}\;{}_i\CA\cr
\lk\b_i\rk={}_i\CN^{-1}{}_i\CD+(\d_{\dc}\;{}_i\CN^{-1})\;{}_i\CA.
\ea\ee
3- Most significant remark is that in the expressions (89, 92, 97), first half of
elements is partial derivative of second half of elements with respect to $\dc$.
This means that ${}_i\CD=\d_{\dc}({}_i\CA)$.

Finaly we can conclude as follows:
\be
\mid n, {\D'}>= \a_{k_1,k_2,\cdots}L_{-1}^{k_1}L_{-2}^{k_2}
L_{-3}^{k_3}\cdots\mid \dc>
+(\d_{\dc}\a_{k_1,k_2,k_3\cdots})L_{-1}^{k_1}L_{-2}^{k_2}L_{-3}^{k_3}
\cdots\mid \D'_p>
\ee
where $\s mk_m=n$.
It seems that appearance of $\d_{\dc}$ has more important rule in logarithmic
operator.

\noindent
{\bf Acknowledgment:}

We would like to acknowledge  A. Aghamohammadi, M. Khorrami,
 M.R. Mohayaee, V. Karimipour and S. Rouhani for
their useful discussions.
The authors are indebted to A. Morozov for his continuous encouragement.

\end{document}